# Algorithm for Back-up and Authentication of Data Stored on Cloud

Manali Raje
Department of Information Technology
Maharashtra Institute of Technology
Pune, India
manali04raje@gmail.com

Debajyoti Mukhopadhyay
Department of Information Technology
Maharashtra Institute of Technology
Pune, India
debajyoti.mukhopadhyay@gmail.com

*Abstract*-Everyday a huge amount of data is generated in Cloud Computing. The maintenance of this electronic data needs some extremely efficient services. There is a need to properly collect this data, check for its authenticity and develop proper backups is needed. The Objective of this paper is to provide Response Server, some solution for the backup of data and its restoration, using the Cloud. The collection of the data is to be done from the client and then the data should be sent to a central location. This process is a platform independent one. The data can then be used as required. The Remote Backup Server facilitates the collection of information from any remote location and provides services to recover the data in case of loss. The authentication of the user is done by using the Asymmetric key algorithm which will in turn leads to the authentication of the data.

## I. INTRODUCTION

The cloud can be said as an on-demand computing, for anyone with a network connection, which provides access to many applications and data, from any device, anytime, anywhere. The consumer-level cloud acts as digital repositories for data and the data from any device which supports internet can be accessed. Cloud Computing allows sharing of resources, information, software which are provided to users on their demand.

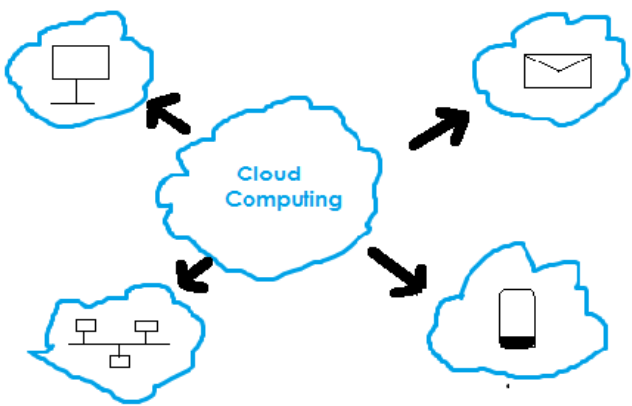

Figure 1. Cloud Architecture

Cloud Computing is the most widely used Technology today which has evolved over time, moving ahead of all the technologies of Computing. The need for cloud's use is increasing day by day as many applications are being deployed on cloud for various numbers of reasons, security and easy access among few. The advantages are many and this is proving to be a boon. Services are offered at a very fast rate and not much effort is needed for that. Storing the data online is one of the best services offered by cloud. The following figure shows the cloud's structure (Figure 1).

Cloud allows the users to make use of its servers to store data. Doing so, does not require the maintenance of onsite resources. This data which is stored can be very sensitive as it belongs to different fields, for example it may be the Medical records of some person or data from the social networks. Privacy and Security become the concern in cloud computing as the remote server obtains the services. The data which is stored online can be accessed by many users at the same time. With this data, there is always a risk of Human fault, equipment failure, network loss etc. Cloud is prone to the Byzantine attacks leading to the failure of the storage systems. Different techniques have been proposed and studied to provide authentication on the data as a measure for data modification problem. Data encryption is one of the ways. Neither the Cloud nor the user should deny of any operations it performed. The aim being that the authentication of the user should be done and also the stored data should not be manipulated .At the same time privacy of the user should be maintained and the user's identity should not be disclosed. The cloud can hold the user responsible for the data it obtains, and the cloud itself is also responsible for the services it provides. The one who stores the data is also validated. The Cloud should only be able to satisfy the query without knowing what the query is. For this, the cloud accesses only the key words and returns the result. The use of public key cryptographic methods allows the authentication of the user . But the Cloud has a changing nature and this can lead to the modification of the data stored. This changing nature is known Data Dynamics. Due to the limitations on Storage and backup of data there is a need for data integrity. It is needed as the Cloud handles huge data.

The recovery of the corrupted data files or the leaked information is an impossible task. So for this reason, there should be some provision to collect the data at some more places and present it to the user in case the data from the main cloud is lost. The storing of data at some place is called as taking a backup and this storage has to be done in a platform independent manner. Also the stored data should be secured in such a way that it will remain safe even in case of any manhandling or theft.

## II. RELATED WORK

Various data backup and recovery techniques are discussed in this section. The techniques used before lacked good performance, were not cost efficient, and compromised with privacy and security. The parity cloud service provides easy recovery of data as compared to any other technique. This technique is based on parity recovery service, which allows the data to be retrieved with a high probability, by using the Exclusive-OR method for the parity information [2].

HSDRT technique handles the laptop and the smart phone users. Encryption using high speed along with the data transfer mechanism for distributed data is used here. This approach has a drawback of high implementation cost and redundancy [1].

The Efficient Routing Grounded on Taxonomy (ERGOT) Technique [3] does the semantic analysis but does not focus on time and complexity of implementation. This approach provides an efficient recovery based on similarity. Another approach known as the Cold and the hot approach increases the cost of implementation as the amount of data increases. The backup is performed as the service fails [4]. Once the service fails only then the backup is made in the cold backup approach. If the services are available then the backup won't be done. The services are in the activated state in Hot Backup.

The authentication literature is as follow. The Storage system may fail due to the byzantine attacks [5]. Encryption using only the keywords is also studied. Only the keywords are searched and returned. There is no need to know the entire query [7]. Public key cryptographic encryption techniques are mentioned in [6].

Retrieval of the data using these techniques is not efficient. The Remote data backup server comes into picture here and is discussed further.

## III. EXISTING APPROACH

The Cloud seems to be making headline everyday with some new innovation coming up every day. Through different interfaces large amount of data is made available to the users. The information from different sources is said to be moved on to the cloud when all of it is collected and stored on cloud. The cloud has two components associated with it, namely the Client and the Service Providers. Having an extra copy of anything or providing a replacement of anything means a backup. So when we say that we are having a Remote Data Backup Server what we mean is that a server is located far from where the actual data is stored. This server has the same feature as that of the main cloud. Central Storage refers to the Main Cloud whereas the Remote Storage refers to the Remote backup.

The Cloud should possess the following features.

- It should provide Integrity to the data which deals with the data's structure and state. It takes care that the data is not mishandled.

- It should provide security to the data stored on the cloud. User and data authentication should be checked thoroughly.

- The Cloud server should be available as and when required by the users. The services should not be denied.

- The users and the cloud should take responsibility of all the operations it performs. There should be no denial in these matters

- There should be consistency of the data which is stored on the cloud. The data stored should support the storage systems.

The Seed Block Algorithm mentioned here is used for the backup purpose. It provides an efficient solution for backup of data. The algorithm is as follows:
Initialization of the Main cloud: M; Remote Server: R; Client: C;

**Files:** z and z'; Seed block: A; Random Number: r; Client ID: C_id

**Input:** The file z created by C and the random number r generated at M.

The output: Recovered file z after being deleted from the main cloud.

**Step 1:** int r=random (); // *Random Number Generated*

**Step 2:** Seed Block A is created for every client on cloud and stored on the remote server.

A= r EXOR C_id.

**Step 3:** z' created if any changes are made to the file stored on Main Cloud, then we have.

z'=z EXOR A

**Step 4:** z' is stored at the Remote Server

**Step 5:** z is deleted because of some reason then z=z' EXOR A is used.

**Step 6:** Return z to C.

**Step 7:** End

The algorithm uses the Exclusive-OR Operation when executed. Consider two files P and Q. When 1 and 2 are

EXORed its result is stored in R. A backup and recovery is provided. A main cloud server, its clients and the remote data server are the major components. Every client has a unique identity associated with it. These are EXORed when registration is done on the cloud. An individual block is created when the client's id and the random number are exored with each other. This individual block is known as the seed block, the seed block is then on the remote server. The storage on the main cloud occurs when a file is created for the first time. While the storage on the main file is taking place, simultaneously the file is EXORed with the seed block and is stored on the remote server. Incase the Cloud crashes or the file is accidentally deleted then it can be retrieved from the exored file which is stored on the Remote server.

The repository, web service, the database and the users are the main components. The client's laptop or the smart phone maintains the application (Figure 2). This application is platform independent and can be mounted on any operating system. The data from these devices which is independent of the platform is then sent to the Central repository. Data is checked for validation and then passed on to the virtual database. The users and the database is connected through web services and the customized data is sent over to the user. Following functions are performed by the various components of the system:
Verification of the user and the information before being stored onto the database.

- Details of the user, updation time is also recorded.
- Handling of multiple requests by the repository.
- Sharing of data amongst users.

Validating the data is done before storing it in the database. Doing this allows the user to be confident that the data is safely stored in the remote back-up server and can be retrieved as and when required.
There are two types of users
- Internal users: Internal users include the administrators or the domain experts.
- External Users: It includes any person who wants any information or data.

The information from both the internal as well as the external user is stored in the database. The records and data as per the requirement are maintained by the admin. The admin also maintains the web services and the cloud server. Manual or automatic restoration can be done. The flow of the work goes as follows.
The following figure shows the flow of the data.

The data processing web server handles multiple requests from different users.
This approach provides robustness, easy to use and proper backup facilities, flexibility, availability, portability, easy maintenance.
*Flexibility:* The system can add on any new feature also supports the features added.

*Portability:* Is platform independent and works efficiently in any environment.
*Fast*: Faster than manually maintained systems. Has low scope for human error.
*User Friendly*: Gives user friendly ways to manage the system along with low cost implementation.
*Reliability*: Reliable, secure. Provides authentication.
*Proper backup facility*: Same sized data is recovered in case off loss or server failure.

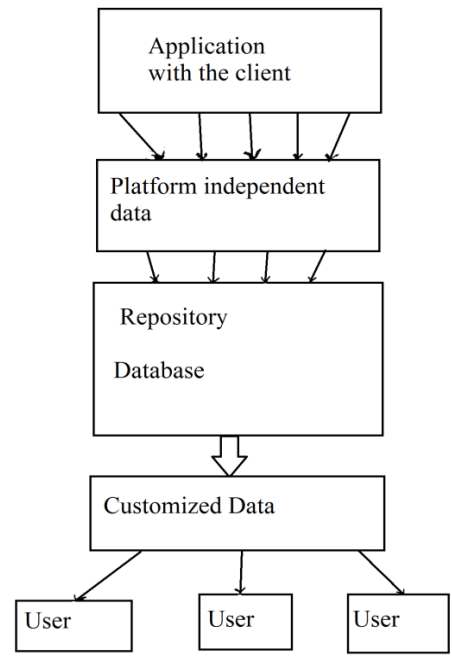

Figure 2. Flowchart

IV. PROPOSED WORK

Now, as the data gets stored on the main cloud and then by using the EXOR technique it is stored on the Remote server, the question of providing security arises. The user of the data should be authenticated which in turn will authenticate the data before storing it on the cloud. For this purpose various kinds of cryptographic techniques can be used. Taking into consideration the advantages and disadvantages of the techniques, encryption using the public and private key is done on the data. The encrypted data is then Exored and stored on the Remote Server. For storage on the Cloud Server the data can be stored in its original form or in the encrypted form. For the proposed approach the work proceeds as follows. There are three components namely the admin, the information seeker and the information provider. The admin assigns the tokens to the seeker as well as the information provider. The seeker or the provider on presenting the token to the admin when asked will then be given the secret key by the admin which can be used for the encryption or the decryption purpose. The admin will be keeping a record off all the members it has provided with the token and the ones involved in the communication. After this check is done the message is encrypted by the provider. The cipher text is generated on encrypting the message and then sent to the remote server by applying the EXOR technique on it. When the main cloud has crashed or the original file has lost from the main cloud, it can be retrieved from the remote server. What the seeker gets is the Exored file which has to be decrypted by using the public and the private key. Again the admin checks its database for all the registered and authorized users.

The encryption by the Information provider is done as follows:

A token @ is collected from the once registration is done, *p* is the signature
*token @ = (u, public key, private key, ρ)*
The admin has the all access to the secret keys
After this token is issued the provider encrypts the information by using its own private key and the public key assigned to it by the admin. Let the public key be a1 and the private key be a2.
*C= Encrypt (msg, a1, a2)*

This information is then Exored and sent to the cloud. The admin checks if all the values are correct and the encryption process has proceeded properly. Record is maintained to authenticate all the users.
After the data is stored on the cloud it is available for the seeker to read it. The cipher text is being presented to the user who then decrypts it using the public and the private keys. Again the admin checks on with the database for the authenticated users and the data is provided on validation. Once the encryption is done the data will be exored and then stored on the remote server. The expected result will be as shown in fig. 3. when the algorithm for backup is applied. Once this part is done it will proceed for the authentication part, which is still under working.

## VI. CONCLUSION

The large amount of data generated on Cloud each and every day can be securely stored using encryption and also can be retrieved in case the files from the main cloud are deleted or lost due to any reasons or even in case of Cloud crash. By doing the encryption we are assuring the safety of the data stored on the cloud.

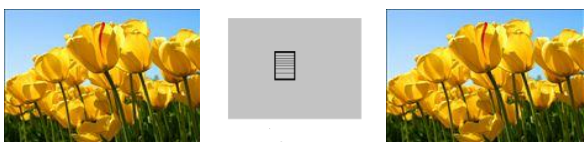

Figure 3. Expected Result

## V. MATHEMATICAL MODEL

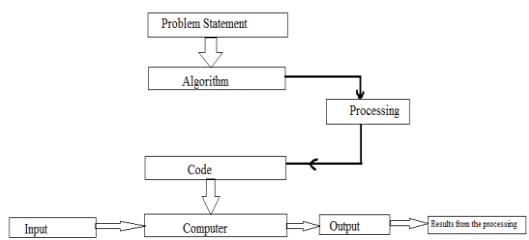

Figure 4. Mathematical Model